# Capillarity Theory for the Fly-Casting Mechanism

Emmanuel Trizac[a], Yaakov Levy[b], and Peter G. Wolynes[c*]

[a]Laboratoire de Physique Théorique et Modéles Statistiques, CNRS UMR 8626, Université Paris-Sud, F-91405 Orsay Cedex, France; [b]Department of Structural Biology, Weizmann Institute of Science, Rehovot 76100, Israel ; and [c]Department of Chemistry and Biochemistry, University of California, San Diego, 9500 Gilman Drive, La Jolla, CA 92093-0371

**\* Corresponding author:** pwolynes@ucsd.edu

**Running title:** "Biophysical criteria for fly-casting binding"

**Keywords:** protein folding, binding mechanism, disordered proteins, fast folding

**Abstract**

Biomolecular folding and function are often coupled. During molecular recognition events, one of the binding partners may transiently or partially unfold, allowing more rapid access to a binding site. We describe a simple model for this flycasting mechanism based on the capillarity approximation and polymer chain statistics. The model shows that flycasting is most effective when the protein unfolding barrier is small and the part of the chain which extends towards the target is relatively rigid. These features are often seen in known examples of flycasting in protein-DNA binding. Simulations of protein-DNA binding based on well-funneled native-topology models with electrostatic forces confirm the trends of the analytical theory.

**Introduction**

It is becoming clear that protein folding and protein functioning are often overlapping processes. For cooperatively folding proteins, a free energy barrier separates the folded configurations from the unfolded ones. If this barrier is sufficiently high, we expect the traditional "function follows folding" paradigm to be valid. Many proteins however are unfolded before they function(1-5). Other proteins have native ensembles separated by only a small barrier from their disordered states. It has recently been argued that in some circumstances there may be no folding barrier at all(6-8). Clearly in all these cases folding and function must be coupled. In addition, partially surmounting the folding barrier, even if it is high, may short circuit the otherwise large barriers that would accompany allosteric changes, if the protein were required to always remain intact. These two ways of coupling the folding landscape and the functional landscape have been termed "fly-casting" (9-11) (for the binding process) and "cracking" (for allosteric change)(12-14). In the fly-casting scenario, a protein may bind from a relatively large distance, thereby enhancing its capture radius: the trade-off is between the entropy cost from extending a subdomain and the energy gained upon binding to a target. While the disordered proteins may have slower translational diffusion comparing to globular proteins, their intrinsic flexibility imposes fewer constraints on binding and therefore they may have faster binding to their binding partners(10, 15). It is possible that the biological function of downhill and ultra-fast folding(16-20) is to achieve fast binding via fly-casting. In this paper, we explore the interplay between the fly-casting and the folding barrier using the capillarity model for protein folding, which in its simplest form ignores many of the structural features of the native protein.

Despite the intrinsic complexity of the folding of a random heteropolymer, most proteins have evolved to fold via a funneled energy landscape(21, 22). This evolved feature greatly simplifies the physics of the folding process because the primary free energy scales involved are then the entropy of organizing the chain into a specific topology and the stabilization energy. These free energies must balance near the physiological folding temperature, giving a single parameter for that determines the specific topology free energy profile. The remaining smaller energy scales come from the ruggedness of the landscape (which is reflected in the rate of conformational diffusion) across the landscape and a cooperativity energy that comes from the nonadditivity of inter-residue forces which has a large component from solvation. The latter effect is very analogous to the surface tension of a liquid condensing from a gas. The capillarity picture of folding takes over the picture of nucleation at a first order transition more or less intact to describe the change from folded to unfolded states(23). This picture might be called the "spherical cow" approximation to folding since it ignores, to zeroth order, the specifics of protein topology and connectivity. This great oversimplification is, however, a positive feature when we wish to understand trends, such as the scaling of folding rate with length etc.

In this paper, we use the capillarity picture to look at fly-casting. The model gives crisp criteria for the conditions under which flycasting can occur. The analysis shows the key role played by the ratio of the binding affinity to the folding barrier in determining whether fly-casting will occur. In general fly-casting is encouraged by high affinities and low barriers. Fly-casting and cracking are thought to play a special role in how transcription factors navigate along DNA and brachiate between DNA chains(24). On the basis of the present theory it turns out then to be quite natural that DNA binding proteins have been found to be among the fastest

folders(10, 17). Rigidity of the protein in the denatured state also encourages fly-casting thus suggesting an increased possible role for residual structure in the denatured state.

In addition to outlining the capillarity model of fly-casting in this paper, we compare the theory with results from a topology based simulation model of the protein-DNA binding process. This analysis confirms the trends expected from the capillarity model.

## Methods

**The capillarity model of fly-casting association**

For our purposes, the protein made up of N residues is split into three regions. The first part with $N_1$ residues is structured (i.e., folded) and in contact with the target (assumed to be a relatively simple geometry, such as an interface, membrane, or DNA). At the other terminus of the protein, we have another structured region of $N_2$ residues. Finally, the fly-casting effect translates into an intermediate region with $N_3 = N-N_1-N_2$ unstructured residues, with end-to-end extension L, and perpendicular to the binding substrate. This intermediate section will be treated as a standard homo-polymer coil under stretching. The $N_1$ section being in direct contact with the target, the protein center-of-mass distance to the substrate is $R=L(N_2+N_3/2)/N$. We implicitly neglect the size of ordered regions (1 and 2) compared to the "arm" extension L, and we emphasize that the conformations under study here impose a contact between the protein and the binding substrate. Under those assumptions, the total free energy is written as

$$F = F_{cap}(N_1) + F_{cap}(N_2) + F_{bind}(N_1) + F_{ent}(N_3) \qquad (1)$$

Both $N_1$ and $N_2$ termini contribute to the free energy through a capillarity term(23, 25)

$$F_{cap}(N_i) = \gamma(N_i^{2/3} - N^{-1/3} N_i) + \tau N_i \qquad (2)$$

where $\tau \propto T - T_f$ denotes the deviation from folding temperature $T_f$. The quantity $\gamma$ is analogous to a surface tension. At $T_f$ it is the single parameter that governs the free energy barrier to completely fold the protein:

$$F_{cap}^{barrier} = \frac{4\gamma N^{2/3}}{27(1 - \tau N^{1/3}/\gamma)^2} \qquad (3)$$

In what follows, we will refer to the free energy barrier for folding as the above quantity, evaluated at folding temperature:

$$B = \frac{4\gamma N^{2/3}}{27} \qquad (4)$$

If $\nu$ denotes the Flory exponent(26) of the intermediate $N_3$ part, the corresponding entropy cost of stretching reads $F_{ent}(N_3)$, up to an irrelevant prefactor

$$F_{ent} \propto kT\varepsilon \left(\frac{L}{aN_3^\nu}\right)^{1/(1-\nu)} \qquad (5)$$

where $a$ denotes the residue size (approximately 0.3 nm), and $kT$ is the thermal energy. Whether we consider a Gaussian chain ($\nu = 1/2$) or a swollen coil with excluded volume ($\nu \approx 0.6$) is immaterial, and we will assume for simplicity that $\nu = 1/2$ in the following. More important is the prefactor $\varepsilon$ that accounts for the rigidity of the chain: it can be viewed as the ratio $a/l_p$ where $l_p$ is the chain persistence length: $\varepsilon = 1$ for a fully flexible chain, whereas we will consider $\varepsilon = 0.2$ for a typical protein. Finally, as far as the binding term is concerned, several models can be considered; in the simplest approximation, we shall take $F_{bind}(N_1) = -\delta N_1 kT$. Several constraints should be enforced, such as $L \leq N_3 a$, or others resulting from working at fixed R.

**Simulation model of biomolecular association**

We study the effect of the folding barrier and binding affinity on the fly-casting characteristics using a coarse-grained simulation model for a system that has been reported to follow the fly-casting mechanism. The association of the Ets protein to its specific DNA binding site was studied using a native topology based model (Go model) supplemented by electrostatic interactions. The native topology based model takes into account only interactions that exist in the native structure and, therefore, includes mainly topological frustration. Adding nonspecific electrostatic interactions contributes energetic frustration between the protein and the DNA as well as within the protein. Purely structure based models have already been used to study the folding of many monomeric proteins that fold in a two-state fashion(27-30) or that fold via a more complicated folding scheme. They have also been used to study higher molecularity reactions such as dimerization and tetramerization(11, 31-33).

In our study, the protein and the DNA are modeled using a reduced representation. Each residue is represented by a single bead centered on its α-carbon ($C_\alpha$) position. Adjacent beads are strung together into a polymer chain by means of a potential encoding bond length and angle constraints. The secondary structure is encoded in the dihedral angle potential and the non-bonded (native contact) potential. In the framework of the model, all native contacts are represented by a 10-12 Lennard Jones form without any discrimination between the various chemical types of interaction. The residues do not have any chemical identity (the information for folding is encoded in the structure) but some have a point charge. Accordingly, positively charged residues (Arg and Lysine) have positive point charge and the negatively charged residue (Asp and Glu) have a negative point charge. The other beads are neutral. Details of the coarse-grained modeling of the protein-DNA system can be found in previous publications(10, 34).

To enhance the sampling of binding events, a constraint is applied on the protein-DNA system so that they are confined to a sphere of radius 40 Å centered at the center of mass of the DNA that is kept frozen. For each electrostatic strength, several constant temperature molecular dynamics simulations were performed starting from an unbound (folded or unfolded) protein conformation. At a given temperature several trajectories were collected starting from different initial conformation and velocities. The length of the simulations was determined by the demand of having several transitions of folding/unfolding or binding/unbinding at the transition temperatures. The multiple trajectories were combined using the Weighted Histogram Analysis Method (WHAM(35)) to calculate thermodynamic properties of the systems.

To investigate the effect of the folding barrier of Ets and its affinity to DNA, the association of Ets with its DNA cognate was studied for different values of dielectric constant. In addition, the folding barrier was tuned by increasing the cooperativity by incorporating a non-additive term to the Hamilitonian(36).

**Results**

**The persistent limit**

The case of a persistent fly-casting "arm" is straightforward to work out and yields an interesting benchmark. In this limit indeed, the ratio $\varepsilon$ is small, and the entropy cost of the arm irrelevant. We therefore assume here that the middle part of the protein is fully stretched, so that $L = N_3 a$, which therefore provides the most favorable conditions for the occurrence of fly-casting. It can be shown that whenever the total free energy $F$ becomes negative, the only folded region of the protein is that part in contact with the membrane ($N_2 = 0$) and $N_1$ takes its maximum value allowed (for a given value of $R$): $N_1 = N - \sqrt{2NR/a}$. Starting from a large value of $R$, $F$ first increases upon decreasing $R$, which corresponds to the energy penalty of folding of $N_1$-like units, then decreases and vanishes at $R=R^*$ with

$$R^* = \frac{1}{2}\left[\left(1+\frac{\delta N^{1/3}}{\gamma}\right)^{-3} - 1\right]^2 \quad (6)$$

For $R<R^*$, $F<0$; of course, the states with positive free energy at $R>R^*$ are metastable, and are obtained enforcing a contact between the protein and the binding substrate for all admissible $R$. As a consequence, in the range $R>R^*$, it is more favorable to approach the substrate in the collapsed state (with $N_1 = N_2 = 0$), with a vanishing free energy and no contact to the substrate. The free energy profile is shown in Fig 2.

Increasing the ratio $\delta/\gamma$ increases $R^*$, but preserves the shape of the curves shown in Fig 2. Although it is thermodynamically favorable to fly-cast when $R<R^*$, kinetic effects may preempt the phenomenon, and it is of interest to compute the associated free energy barrier. To this end, we notice that as far as the folding of $N_1$ residues is concerned, a folding temperature mismatch $\tau$ without binding is equivalent to a (negative) binding term at the folding temperature; taking advantage of this $\delta \leftrightarrow -\tau$ correspondence, Equation (3) yields an energy barrier

$$F^{barrier}_{\varepsilon=0} = \frac{4\gamma N^{2/3}}{27(1+\delta N^{1/3}/\gamma)^2} < B \qquad (7)$$

If the quantity on the left hand side above exceeds some threshold $F^{thres}$, fly-casting will not take place. More explicitly, such a situation of kinetic hindrance occurs for $F^{thres} < B$ under the condition that for

$$\delta NkT < \frac{27}{4} B \left( \sqrt{\frac{B}{F^{thres}}} - 1 \right) \qquad (8)$$

In the remainder, we will refer to $\delta NkT$ as the affinity.

**The general case ($\varepsilon \neq 0$)**

For a given value of $R$, the free energy in (1) depends on 2 variables (e.g., $N_3$ and $L$), while $N_2$ follows from $R = L(N_2 - N_3/2)/N$ and $N_1 = N - N_2 - N_3$. At variance with the persistent limit where the "arm" extension $L$ can take its maximum value ($N_3 a$) at no entropy cost, $L$ is here a variable that follows from minimizing the total free energy. In the relevant parameter range, at fixed $R$ and $L$, $F(N_3)$ is a concave function, which therefore reaches its minimum at the boundaries of the accessible $N_3$ domain. This simplifies the analysis when it comes to finding the optimal $L$ for a given center-of-mass distance $R$. Once the resulting function $L(R)$ is known, we can compute the fly casting distance $R^*$ for which $F$ becomes negative.

In Fig 3, the critical fly casting distance corresponding to the onset of negative free energies is $R^* \approx 0.042 Na$ while the corresponding value of $N_1$ is $N_1^* \approx 0.60$. Increasing the barrier for folding B from 5kT to 10 kT significantly reduces $R^*$ to 0.022 (see also Figure 5), with $N_1^* \approx 0.75$. On the other hand, decreasing the rigidity parameter ε from 0.2 to 0.1 facilitates fly casting ($R^* \approx 0.06 Na$ with nevertheless $N_1^* \approx 0.60$, almost unaffected), keeping γ and δ as in Fig 3.

It turns out that the states that minimize the free energy for $R < R^*$ are such that $N_3$ (number of residues in the "arm", see Fig. 1) takes its maximum value, and that in addition, $N_2$ vanishes. This latter fact allows one to easily relate the temperature dependence in our model (all figures shown above are at the folding temperature $T = T_f$) to its δ –dependence. Decreasing T is equivalent to increasing δ; more precisely, a given affinity parameter δ at $T \neq T_f$ (i.e $\tau \neq 0$) is equivalent to an affinity $\delta - \tau/(kT)$ at $T = T_f$.

Of course, several variants of the present model can be put forward, to refine the analysis. Our goal here has been to devise a minimal framework. We have nevertheless checked that the present phenomenology is unaffected by a modification of the binding energy function, so that its $N_1$ dependence saturates beyond a prescribed threshold.

**Comparison to simulations**

We examine the effect of the folding kinetics and of the protein affinity to the target on the existence of fly-casting mechanism using coarse-grained modeling of association of the Ets protein to its specific DNA binding site (Fig. 6a). We have previously shown that the Ets protein binds its cognate DNA sequence via a fly-casting mechanism(10). It was shown that the protein flexibility significantly enhances binding to the DNA comparing to a scenario when the protein binds the DNA as a rigid molecule. The fly-casting can be probed by plotting the free energy as a function of the separation distance between the protein and DNA (Fig. 6b). These plots indicated a coupling between folding and protein-DNA assembly, even for weak electrostatic forces, resulting in a change of the capture radius of the target. A sharper decrease of the free energy curve is observed when flexibility and electrostatics are both included in contrast to when electrostatic forces alone guide a rigid protein.

To examine the effect of the folding barrier and affinity to the DNA on the mechanism of protein-DNA association and in particularly the onset of fly-casting, we study the assembly of the Ets protein with its cognate DNA at various dielectric constants and various folding cooperativity(36). These two parameters affect the height of the free energy folding barrier and the free energy of the bound complex. Increasing the folding barrier by incorporating a non-additive term in the protein Hamiltonian results in a milder fly-casting effect as reflected by the milder slope of the free energy when plotted along the separation distance of protein-DNA (Fig. 6B). Accordingly, as the folding barrier increases, fly-casting takes place only with closer proximity of the protein to the DNA. We probe the fly-casting by comparing the distance R** of the free energy curves of protein-DNA assembly at a given energy value (Fig. 6B). We point out that R* and R** are defined differently, yet both distances indicate stronger fly-casting effect as they are larger.

Figure 6C illustrates that R** gets smaller as the folding barrier increases for different values of dielectric constants. This observation is in harmony with the capillarity theory prediction shown in Figures 4 and 5a. The coarse-grained model agrees with the analytical model with respect to the effect of the protein-DNA affinity on the fly-casting. R** is larger for high affinity protein-DNA interactions (Fig. 6D), similarly to the trend shown in Figure 4. We note that it is not practical to make a one to one comparison between the analytical and simulation models due to differences in the two models and their parameterization (e.g., in the analytical model the substrate attraction to the target is short range, while it is longer range in the simulations due to electrostatic interaction). Nonetheless, despite the differences between the two models, they concur in the unraveling of the key biophysical parameters that can affect the binding via the fly-casting mechanism.

**Conclusions**

The conceptual separation often made between folding processes and functional dynamics relies on the idea that there is clear distinction in thermodynamic terms between the two parts of the energy landscape--the denatured and native ensembles. In this paper we have put mathematical flesh on this notion by using a simple model. The capillarity model of folding was generalized to study the folding-binding process. Criteria describing when transient unfolding dominates the binding mechanism can be explicitly found in this model. The extent of flycasting is shown to depend on the ratio of the binding affinity to unfolding barrier height.

Flycasting is also shown to be encouraged by low chain entropy for the unfolded region reducing the free energy cost of reaching far towards the target. These trends are confirmed in simulations of protein-DNA binding using a well funneled, structure based model with structural detail. The predicted criteria for flycasting should be susceptible to test by appropriate protein engineering studies of folding/binding kinetics(37).

**Figure legends**

**FIGURE 1:** A schematic view of a protein that interacts with a target (illustrated by the flat surface) via the fly-casting mechanism. The protein is composed of 3 regions: two folded regions, L1 and L2, and a flexible linker of length L3.

**FIGURE 2:** Persistent limit $\varepsilon = 0$

a) $F/B$ as a function of center of mass distance R to the substrate. Here, B denotes the barrier for folding defined in (4), i.e., in the absence of binding; We took B=5 kT, N=100, $\delta$=0.1 (which corresponds to a binding energy of -10 kT). The value of R where F vanishes defines $R^*$ (see the arrow) and discriminates the states with positive free energies at $R>R^*$ (shown with a dashed line) from the stable ones with $R<R^*$. Consequently, the most stable configuration when $R>R^*$ is for a protein "detached" from the substrate, which can be either a disordered state or the folded state (assuming we are at the folding temperature). Hence, the optimal free energy profile is the one shown by the thick continuous line. For the parameters chosen, Eq. (6) gives $R^* = 0.15\ Na$, which is indeed observed in the figure. Taking for concreteness a residue size of 0.3 nm, this corresponds to $R^* = 4.5$ nm. b) Plot of the number of residues in the region in contact with the substrate, as a function of $R$, for the same parameters as in a). Metastable states are again shown by the dashed lines while the stable states correspond to the thick continuous curve.

**Figure 3.** Free energy profile versus R for the same parameters as in Fig 2, but with $\varepsilon = 0.2$ instead of 0. The inset shows the corresponding value of $N_1$. Metastable states are not displayed.

**Figure 4.** iso $R^*$ contour lines in (barrier B versus Affinity) plot.

**Figure 5.** Cuts through the contour plots of Fig 4. a). Shows the critical fly-casting distance $R^*$ as a function of barrier for folding $B$ in the main graph (at constant binding energy $\delta N$ kT = 10 kT). Inset: same, as a function of binding energy, at constant barrier B = 5 kT). Here, the rigidity parameter is $\varepsilon = 0.2$. b). Shows $N_1^*$ as a function of barrier for folding $B$.

**Figure 6.** Simulation of fly-casting binding in protein-DNA interactions. The binding of the Ets protein to its cognate DNA was shown by native topology based model to follow fly-casting mechanism. A). The crystal structure of the specific complex of the Ets protein with its cognate DNA (pdb code 1BC8). B). Free energy curves as a function of the separation distance between protein and DNA for simulation with different degree of non-additivity. Larger non-additivity results with higher folding barrier. The stars correspond to the R** distances that compare the fly-casting strength. C). The R** as a function of the free energy height of the folding barrier for different values of dielectric constant of the Coulomb interactions. The barrier height was modulated by the non-additivity term(36). d). The R** as a function of binding affinity of the Ets to its cognate DNA.

**References:**

1. **Wright, P. E., Dyson, H. J. (1999) Intrinsically unstructured proteins: re-assessing the protein structure-function paradigm** *J Mol Biol* **293:321-331.**
2. **Dyson, H. J., Wright, P. E. (2005) Intrinsically unstructured proteins and their functions** *Nature reviews* **6:197-208.**
3. **Dunker, A. K., Lawson, J. D., Brown, C. J., Williams, R. M., Romero, P., Oh, J. S., Oldfield, C. J., Campen, A. M., Ratliff, C. M., Hipps, K. W.,** *et al.* **(2001) Intrinsically disordered protein** *Journal of Molecular Graphics and Modelling* **19:26-59.**
4. **Wright, P. E., Dyson, H. J. (2009) Linking folding and binding** *Current opinion in structural biology* **19:31-38.**
5. **Tompa, P. (2002) Intrinsically unstructured proteins** *Trends Biochem Sci* **27:527-533.**
6. **Lia, P., Oliva, F. Y., Naganathan, A. N., Munoz, V. (2009) Dynamics of one-state downhill protein folding** *P Natl Acad Sci USA* **106:103-108.**
7. **Huang, F., Sato, S., Sharpe, T. D., Ying, L. M., Fersht, A. R. (2007) Distinguishing between cooperative and unimodal downhill protein folding** *P Natl Acad Sci USA* **104:123-127.**
8. **Cho, S. S., Weinkam, P., Wolynes, P. G. (2008) Origins of barriers and barrierless folding in BBL** *Proc Natl Acad Sci U S A* **105:118-123.**

9. Shoemaker, B. A., Portman, J. J., Wolynes, P. G. (2000) Speeding molecular recognition by using the folding funnel: The fly-casting mechanism *P Natl Acad Sci USA* **97**:8868-+.
10. Levy, Y., Onuchic, J. N., Wolynes, P. G. (2007) Fly-casting in protein-DNA binding: frustration between protein folding and electrostatics facilitates target recognition *J Am Chem Soc* **129**:738-739.
11. Levy, Y., Wolynes, P. G., Onuchic, J. N. (2004) Protein topology determines binding mechanism *P Natl Acad Sci USA* **101**:511-516.
12. Miyashita, O., Onuchic, J. N., Wolynes, P. G. (2003) Nonlinear elasticity, proteinquakes, and the energy landscapes of functional transitions in proteins *P Natl Acad Sci USA* **100**:12570-12575.
13. Miyashita, O., Wolynes, P. G., Onuchic, J. N. (2005) Simple energy landscape model for the kinetics of functional transitions in proteins *J Phys Chem B* **109**:1959-1969.
14. Whitford, P. C., Miyashita, O., Levy, Y., Onuchic, J. N. (2007) Conformational transitions of adenylate kinase: Switching by cracking *Journal of Molecular Biology* **366**:1661-1671.
15. Huang, Y., Liu, Z. (2009) Kinetic advantage of intrinsically disordered proteins in coupled folding-binding process: a critical assessment of the "fly-casting" mechanism *J Mol Biol* **393**:1143-1159.
16. Kubelka, J., Hofrichter, J., Eaton, W. A. (2004) The protein folding 'speed limit' *Current opinion in structural biology* **14**:76-88.
17. Yang, W., Gruebele, M. (2004) Folding of l-repressor at its speed limit *biophysical Journal* **87**:596-608.
18. Ferguson, N., Johnson, C. M., Macias, M., Oschkinat, H., Fersht, A. (2001) Ultrafast folding of WW domains without structured aromatic clusters in the denatured state *Proc Natl Acad Sci U S A* **98**:13002-13007.
19. Buscaglia, M., Lapidus, L. J., Eaton, W. A., Hofrichter, J. (2006) Effects of denaturants on the dynamics of loop formation in polypeptides *Biophysical Journal* **91**:276-288.
20. Hannes, N, M, J. C., R, F. A. (2009) Direct observation of ultrafast folding and denatured state dynamics in single protein molecules *Proc Natl Acad Sci USA* **106**:18569-18574.
21. Onuchic, J. N., Luthey-Schulten, Z., Wolynes, P. G. (1997) Theory of protein folding: the energy landscape perspective *Annu Rev Phys Chem* **48**:545-600.
22. Onuchic, J. N., Wolynes, P. G. (2004) Theory of protein folding *Current opinion in structural biology* **14**:70-75.
23. Wolynes, P. G. (1997) Folding funnels and energy landscapes of larger proteins within the capillarity approximation *P Natl Acad Sci USA* **94**:6170-6175.
24. Vuzman, D., Azia, A., Levy, Y. (In press) Searching DNA via a "monkey bar" mechanism: the significance of disordered tails *J Mol Biol*.
25. Finkelstein, A., Badretdinov, A. (1997) Rate of protein folding near the point of thermodynamic equilibrium between the coil and the most stable chain fold *Fold Des* **2**:115-121.
26. Gennes, P.G. de ed. (1979) *Scaling concepts in polymer physics,* (Cornell University Press, Ithaca, NY).

27. **Clementi, C., Nymeyer, H., Onuchic, J. N. (2000) Topological and energetic factors: what determines the structural details of the transition state ensemble and "en-route" intermediates for protein folding? An investigation for small globular proteins** *J Mol Biol* **298:937-953.**
28. **Chavez, L. L., Onuchic, J. N., Clementi, C. (2004) Quantifying the roughness on the free energy landscape: entropic bottlenecks and protein folding rates** *J Am Chem Soc* **126:8426-8432.**
29. **Clementi, C., Garcia, A. E., Onuchic, J. N. (2003) Interplay among tertiary contacts, secondary structure formation and side-chain packing in the protein folding mechanism all-atom representation study of protein L.** *J Mol Biol* **326:879-890.**
30. **Finke, J. M., Onuchic, J. N. (2005) Equilibrium and kinetic folding pathways of a TIM barrel with a funneled energy landscape** *Biophysical Journal* **89:488-505.**
31. **Levy, Y., Cho, S. S., Onuchic, J. N., Wolynes, P. G. (2005) A survey of flexible protein binding mechanisms and their transition states using native topology based energy landscapes** *J Mol Biol* **346:1121-1145.**
32. **Levy, Y., Onuchic, J. (2006) Mechanisms of protein assembly: lessons from minimalist models** *Acc Chem Res* **39:284-290.**
33. **Levy, Y., Caflisch, A., Onuchic, J. N., Wolynes, P. G. (2004) The folding and dimerization of HIV-1 protease: evidence for a stable monomer from simulations** *J Mol Biol* **340:67-79.**
34. **Toth-Petroczy, A., Simon, I., Fuxreiter, M., Levy, Y. (2009) Disordered Tails of Homeodomains Facilitate DNA Recognition by Providing a Trade-Off between Folding and Specific Binding** *Journal of the American Chemical Society* **131:15084-+.**
35. **Kumar, S., Bouzida, D., Swendsen, R. H., Kollman, P. A., Rosenberg, J. M. (1992) The Weighted Histogram Analysis Method for Free-Energy Calculations on Biomolecules .1. The Method** *Journal of Computational Chemistry* **13:1011-1021.**
36. **Ejtehadi, M. R., Avall, S. P., Plotkin, S. S. (2004) Three-body Interactions Improve the Prediction of Rate and Mechanism in Protein Folding Models** *Proc Natl Acad Sci USA* **101:15088-15093.**
37. **Ferreiro, D. U., Sanchez, I. E., de Prat Gay, G. (2008) Transition state for protein-DNA recognition** *Proc Natl Acad Sci U S A* **105:10797-10802.**

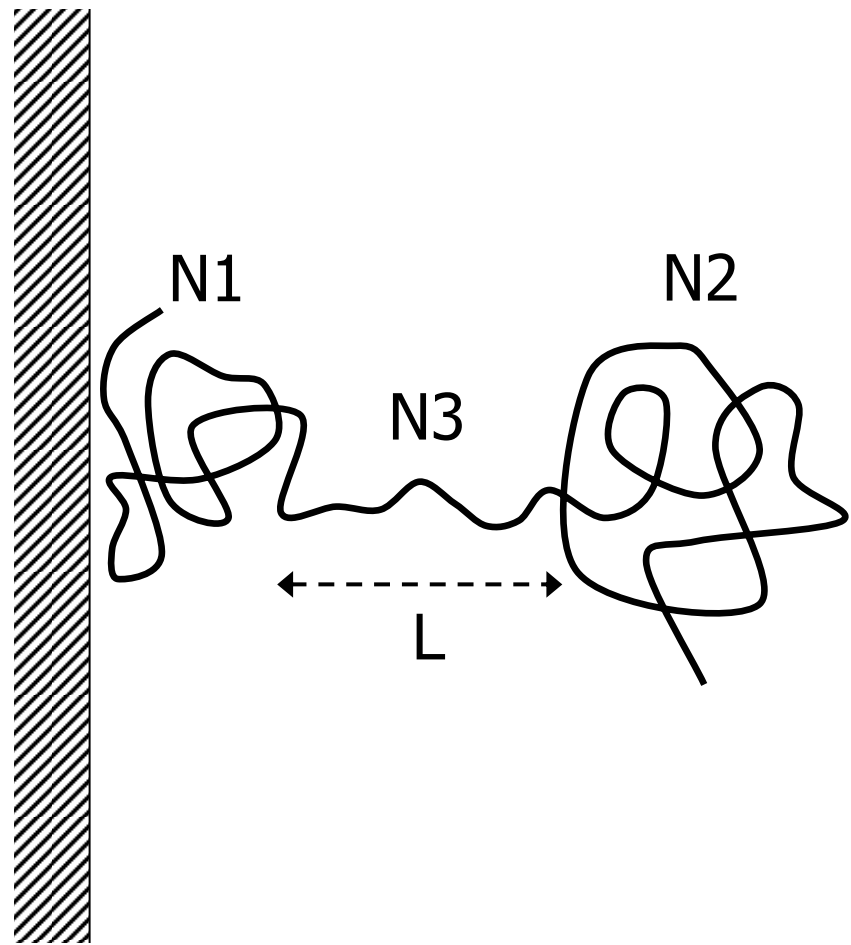

Figure 1

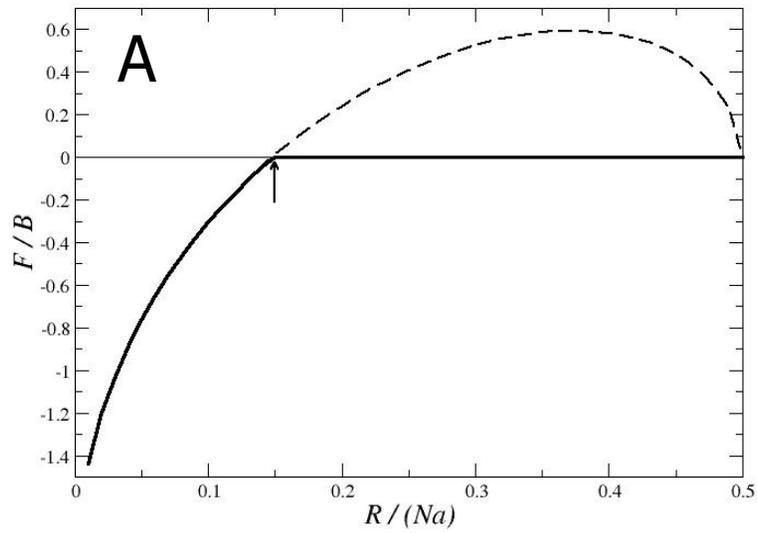 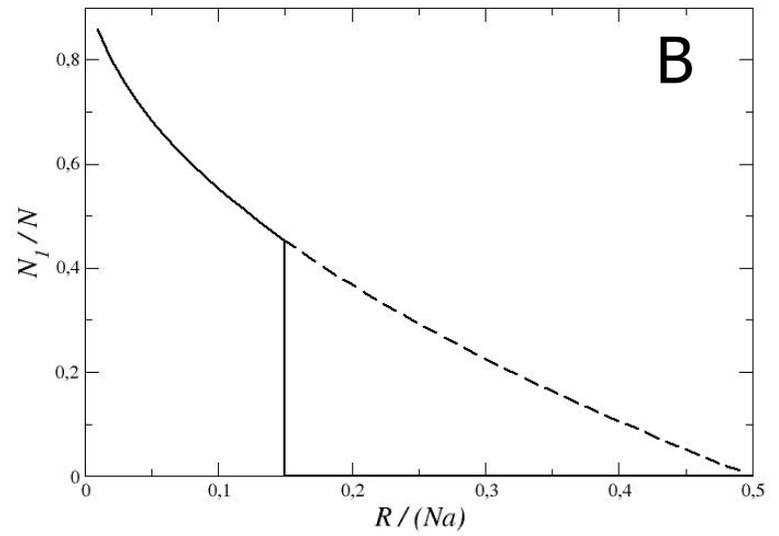

Figure 2

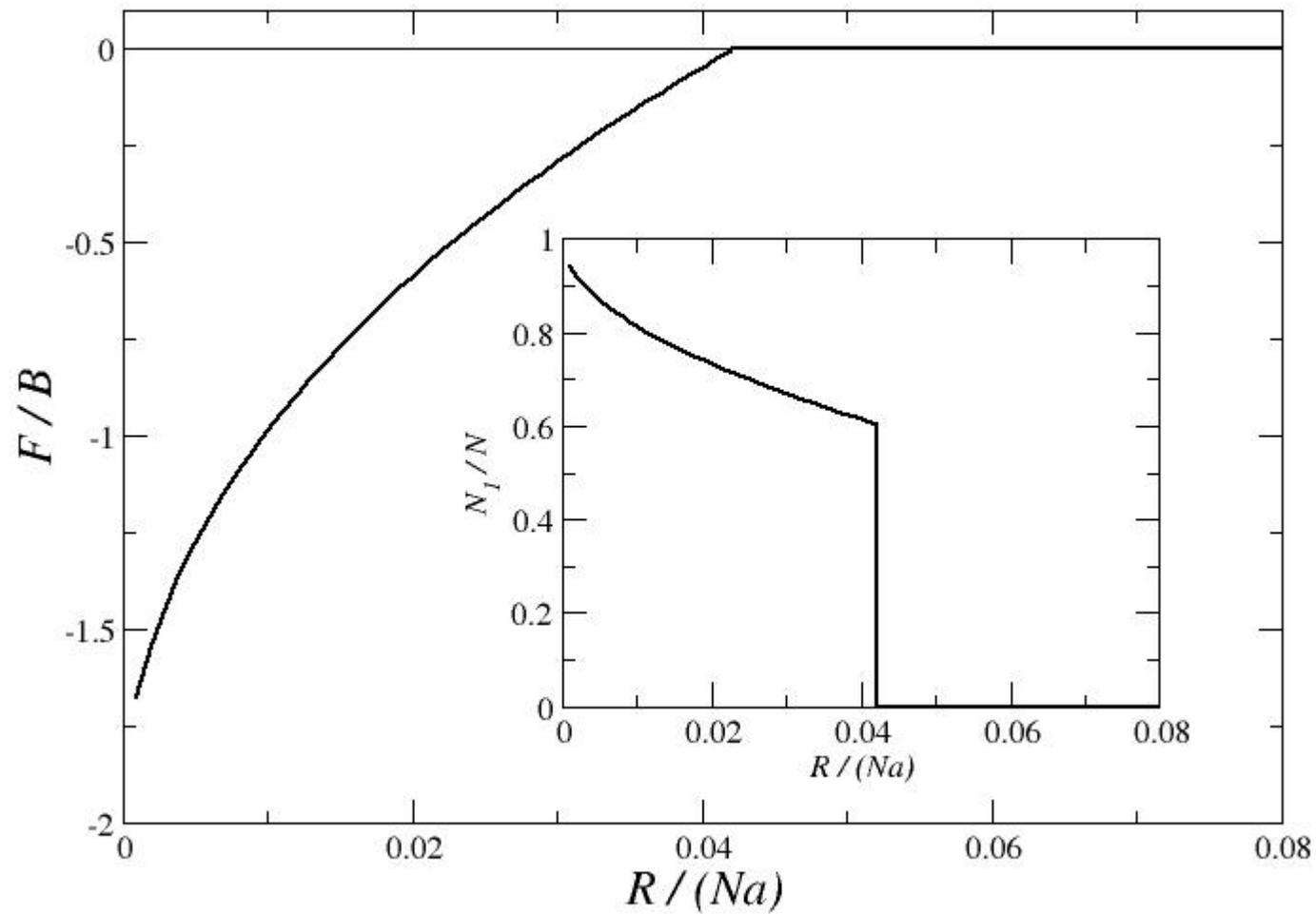

Figure 3

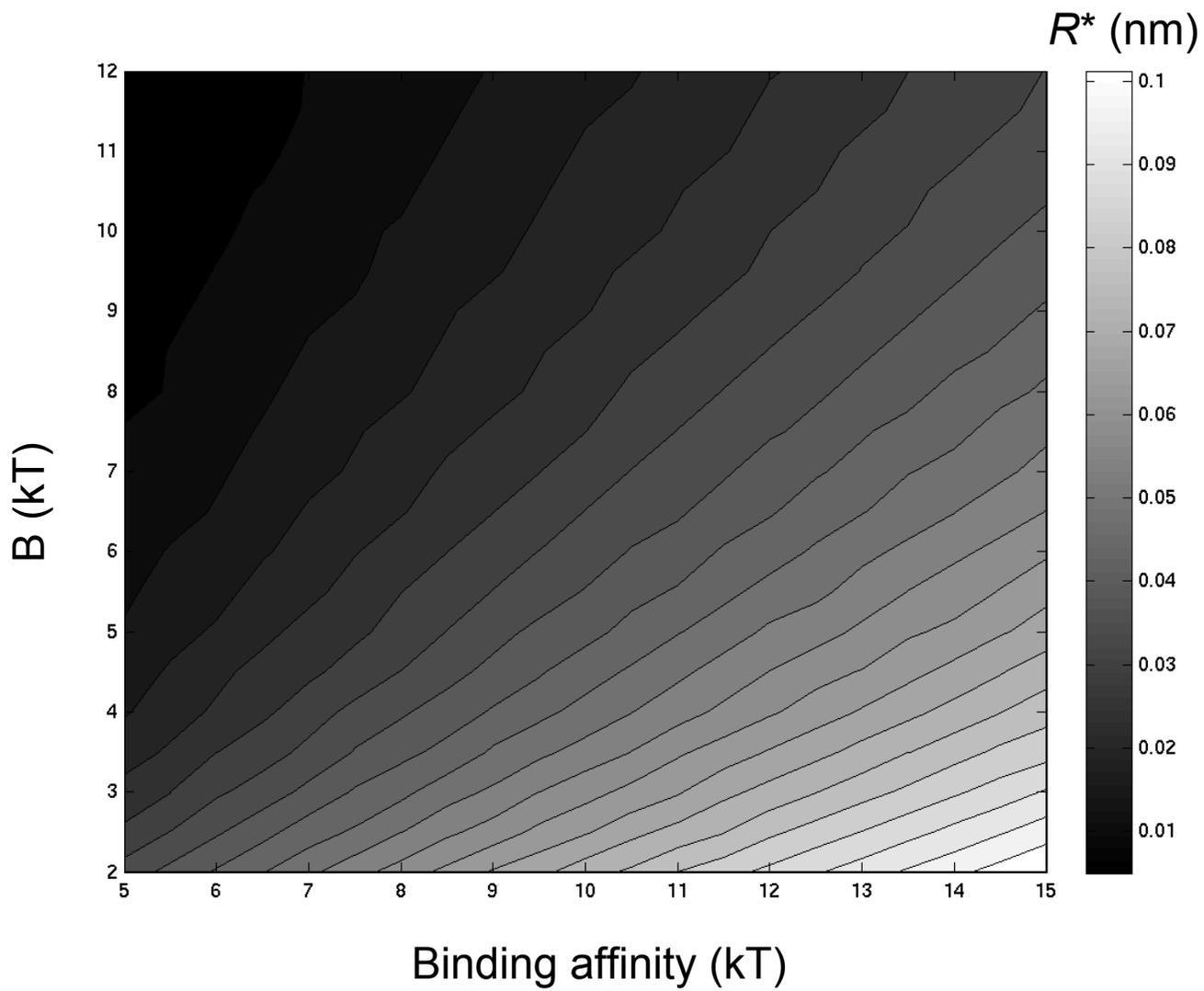

Figure 4

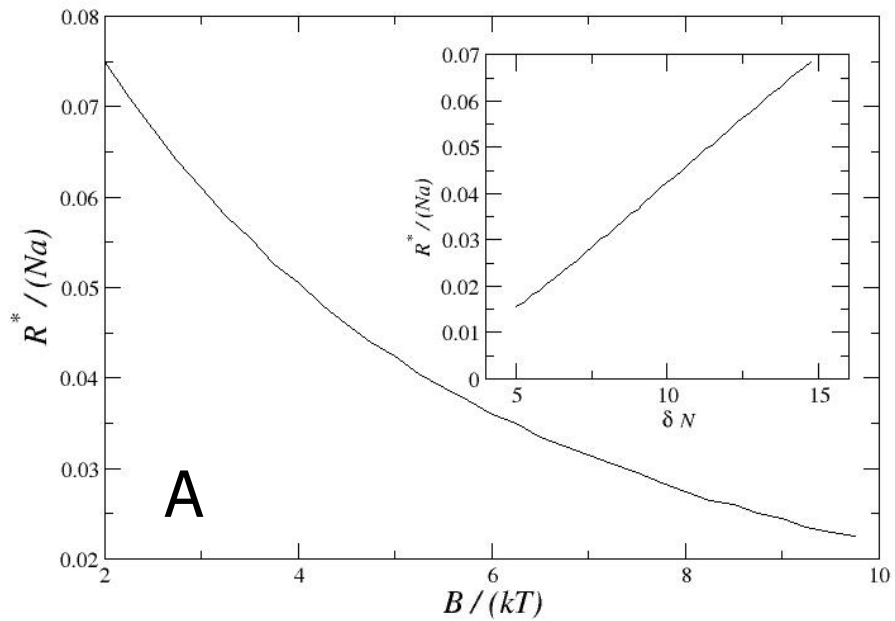

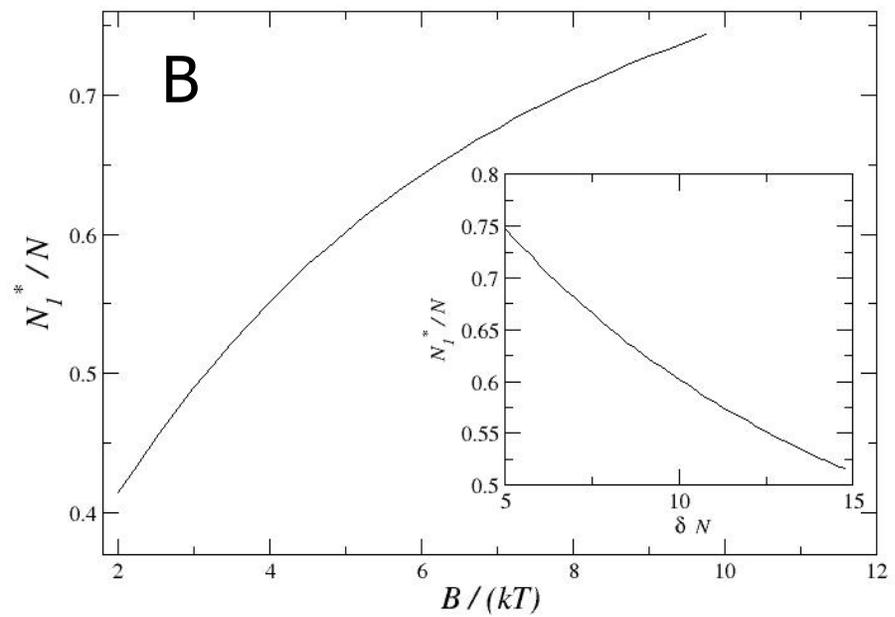

Figure 5

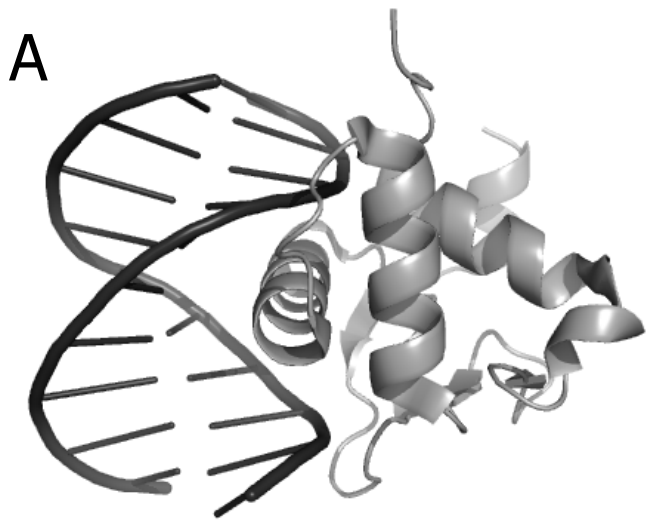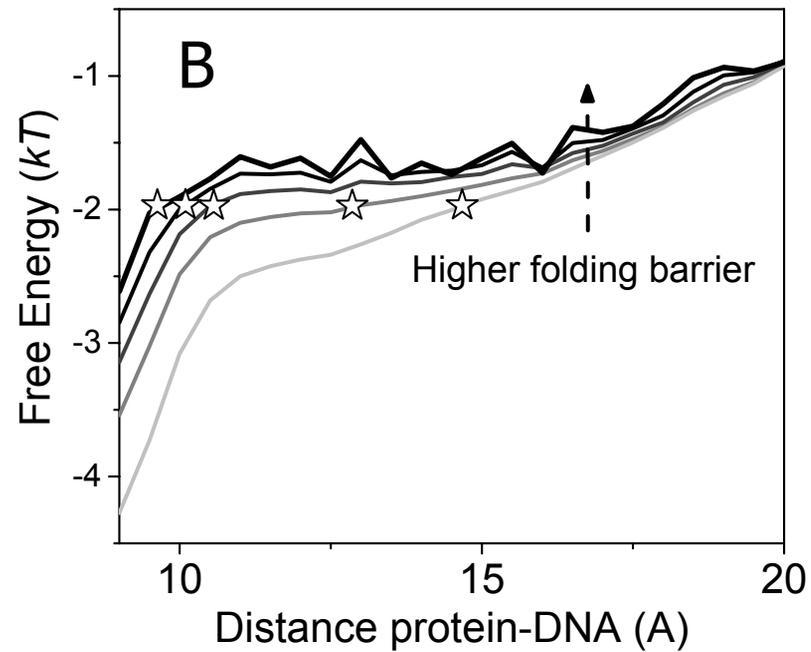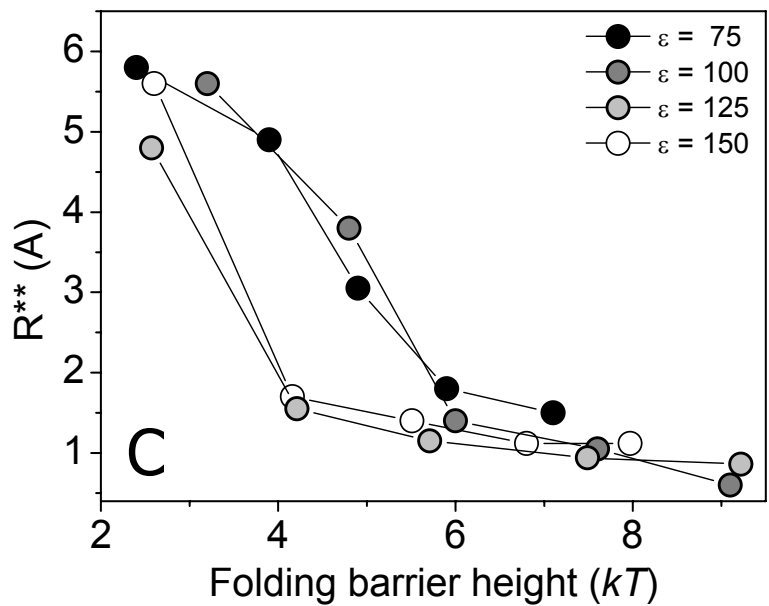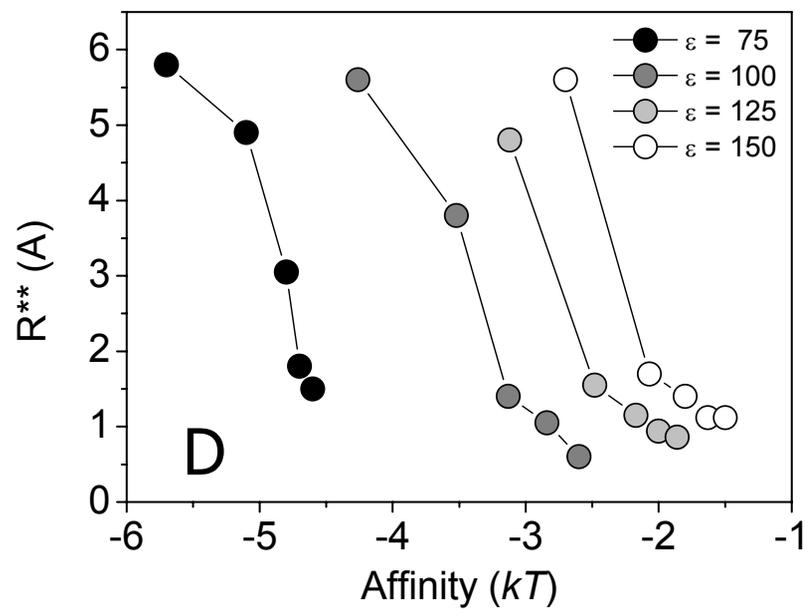

Figure 6